\documentclass[aps,pre,twocolumn,groupedaddress,showpacs,amssymb,amsmath]{revtex4-1}
\usepackage{graphicx}
\usepackage{epsfig}
\usepackage{epstopdf}

\usepackage[T1]{fontenc}
\usepackage{lmodern}
\usepackage{color}
\definecolor{dblue}{rgb}{0,0,0.75}
\definecolor{dred}{rgb}{0.6,0,0}
\definecolor{dgreen}{rgb}{0,0.5,0}

\begin{document}
	
\title{Kinetic Ising models with various single-spin flip dynamics on quenched and annealed random regular graphs.}
\author{Arkadiusz J\k{e}drzejewski, Anna Chmiel, Katarzyna Sznajd-Weron} 
\affiliation{Department of Theoretical Physics, Wroc\l{}aw University of Science and Technology, Wroc\l{}aw, Poland}
\date{\today}

\begin{abstract}
We investigate a kinetic Ising model with several single-spin flip dynamics (including Metropolis and heat-bath) on quenched and annealed random regular graphs. As expected, on the quenched structures all proposed algorithms reproduce the same results since the conditions for the detailed balance and the Boltzmann distribution in an equilibrium are satisfied. However, on the annealed graphs situation is far less clear --  the network annealing disturbs the equilibrium moving the system away from it. Consequently, distinct dynamics lead to different steady states. We show that some algorithms are more resistant to the annealed disorder, which causes only small quantitative changes in the model behavior.  On the other hand, there are dynamics for which the influence of annealing on the system is significant, and qualitative changes arise like switching the type of phase transition from continuous to discontinuous one. We try to identify features of the proposed dynamics which are responsible for the above phenomenon. 
\end{abstract}
\maketitle


\section{Introduction}
Recently a puzzling phenomenon related to the phase transitions has been found in the modified kinetic Ising model, the $q$-neighbor Ising model, with the Metropolis algorithm on the complete graph \cite{Jed:Chm:Szn:15}. Within this model, each spin  interacts only with $q$ spins randomly chosen from its neighborhood. It has occurred that only for $q=3$ there is a continuous order-disorder phase transition, whereas for $q \ge 4$ discontinuous phase transition appears. Moreover, for $q \ge 4$ the hysteresis exhibits oscillatory behavior, i.e., expands for the even values of $q$ and shrinks for the odd values of $q$. Despite the fact that most of the results presented in \cite{Jed:Chm:Szn:15} were derived from the master equation and not only from the computer simulations, the intuitive understanding of the non-monotonicity of hysteresis is still missing. 

Above results are even more confusing if we realize that the $q$-neighbor Ising model on the complete graph seems to be identical with the classical Ising model with the nearest-neighbors (nn) interactions on the annealed random $q$-regular graph ($q$-RRG) \cite{Lee:09,Bol:01,Par:Hoh:17}. The confusion comes from the fact that the Ising model with ferromagnetic nn interactions without an external field has been already examined on a number of different quenched and annealed networks, and it always manifests continuous phase transitions \cite{Pek:01,Agata:02,Dor:Gol:Men:02,Herr:15,Lee:09,Mal:Val:Das:14,Lip:Gon:Lip:15}. Discontinuous or mixed-order phase transitions have been observed only if the range of interactions were infinite \cite{Bar:Muk:14a,Bar:Muk:14b} or on the coupled networks \cite{Such:Hol:09}. Extensive studies have shown not only that phase transition in the Ising model is continuous but also characterized with the same set of bulk critical exponents on quenched and annealed networks \cite{Lee:09}.  

Only very recently it has been shown that  the $q$-neighbor Ising model is a limiting case of a non-equilibrium system with two 
heat baths: the Ising spins are in thermal contact with the heat bath $B_S$ with temperature $T_s$, whereas links are in thermal contact with another bath $B_L$ with temperature $T_L$ \cite{Par:Hoh:17}. For $T_L=\infty$ such a generalized model reduces to the $q$-neighbor Ising model investigated in \cite{Jed:Chm:Szn:15}. It has been also shown that for  $T_L=\infty$ there is nonzero positive heat flux, which confirms that the $q$-neighbor Ising model is indeed out of equilibrium.

Paper \cite{Par:Hoh:17} explains the puzzle of seemingly contradictory results -- continuous phase transition in the equilibrium Ising model on the annealed network \cite{Lee:09} vs. discontinuous phase transition in the $q$-neighbor Ising model \cite{Jed:Chm:Szn:15}. However, many interesting questions are still open. In particular it is still unclear if   
we can fully explain the switch from continuous to discontinuous phase transition on the basis of non-equilibrium regime. Is it possible that other dynamics, even for $T_L=\infty$ (out-of-equilibrium), would lead to the same results as the equilibrium model? It seems quite probable, because results presented in \cite{Lip:Gon:Lip:15} suggest that for the heat-bath dynamics the transition would be continuous, at least on the annealed directed graph. Summarizing, in this paper we ask the question about the role of dynamics in the non-equilibrium kinetic Ising model. Referring to this question we examine the kinetic Ising model with several dynamics (including Metropolis and heat-bath) on quenched and annealed  $q$-RRGs. Precisely, we study several different algorithms (dynamics) that belong to a broad class of a single-spin-flip dynamics (i.e., Glauber dynamics, in a broad sense) \cite{Glauber,God:Luc:05}. Within such dynamics each spin is flipped ($S_i \rightarrow -S_i$) with a rate $W(\Delta E)$ per unit time, and this rate is assumed to depend only on the energy difference implied in the flip:
\begin{equation}
\Delta E=E_{\text{new}}-E_{\text{old}}=2J\sum_{\langle i,j\rangle} S_iS_j.
\label{eq:dE}
\end{equation}
We investigate only such transition probabilities that on the quenched graphs fulfill detailed balance and lead to the Boltzmann distribution in a stationary state, which means that they satisfy the following general condition:
\begin{equation}
\frac{W_{\text{old} \rightarrow \text{new}}}{W_{\text{new} \rightarrow \text{old}}}=\exp(-\beta \Delta E ).
\label{det_bal}
\end{equation}
This is quite obvious that on the quenched graph they should all lead to the same result, which will be checked in the next section. However, on the annealed graph the situation is far from being obvious, and we will show that the results strongly depend on the dynamics. The most popular choices for $W(\Delta E)$ are Metropolis \cite{Met:etal:53} and heat-bath dynamics \cite{Jan:08}. For the Metropolis transition probability:
\begin{equation}
W_\text{M}=\min[1,\exp(-\beta \Delta E)],
\label{rate_M}
\end{equation}
and for the heat-bath:
\begin{equation}
W_{\text{HB}}=\frac{1}{1+\exp(\beta \Delta E)}.
\label{rate_HB}
\end{equation}
However, there are many other possibilities that fulfill the required condition given by Eq. (\ref{det_bal}), and several examples will be investigated later in this paper.

\section{The model}
 In the previous paper \cite{Jed:Chm:Szn:15} we have investigated the $q$-neighbor Ising model,  which was basically the Ising model with the Metropolis dynamics \cite{Met:etal:53} with seemingly small modification that was inspired by the $q$-voter model \cite{Cas:Mun:Pas:09,Nyc:Szn:Cis:12,Mob:15,Mel:Mob:Zia:17}. Within the $q$-neighbor Ising model we consider a graph that consists of $N$ nodes, and each node is occupied by exactly one spin, described by a dynamical binary variable $S_i = \pm 1, i=1,...,N$. In each elementary update a single, picked at random, spin $S_i$ interacts with its $q$ neighbors, also randomly chosen from its entire neighborhood. Generally, two methods of selecting neighbors are possible -- with or without repetitions. For example, in the original $q$-voter model repetitions were possible, i.e., a given neighbor could be selected more than once \cite{Cas:Mun:Pas:09}. However, later also the $q$-voter model without repetitions has been investigated \cite{Nyc:Szn:Cis:12}. The version without repetitions  is of course suitable only for the selected topologies in which the minimal degree of a node $k_{min} \ge q$. For the complete graph both methods are applicable for arbitrary value of $q < N$ and give roughly the same results. 
Here, we continue the study on the $q$-neighbor Ising model started in \cite{Jed:Chm:Szn:15}; therefore, we use the same method as previously, i.e., without repetitions. 

It is worth to notice that for the version without repetitions, that we use here, instead of defining the $q$-neighbor Ising model on the complete graph, one could think about the Ising model on the annealed random $q$-regular graph ($q$-RRG) \cite{Bol:01}. Degree distribution for such a graph is fixed and equals $P(k)=\delta_{kq}$, but the arrangement of links between nodes is random and changes in each update.  Within such a reformulation, the algorithm of a~ single update consists of 3 consecutive steps: 
\begin{enumerate}
	\item Randomly choose a spin $S_i$
	\item Calculate the change of the energy involved in the flip of $i$-th spin given by Eq.~(\ref{eq:dE})
	\item Flip the $i$-th spin with probability $W(\Delta E)$
\end{enumerate}
 In the previous paper, the case with  $W(\Delta E)=W_\text{M}$ given by Eq. (\ref{rate_M}) was investigated \cite{Jed:Chm:Szn:15}. In this paper we will use several different dynamics including heat-bath given by Eq. (\ref{rate_HB}). Beyond this generalization of
 $W(\Delta E)$, formally exactly the same algorithm was used in \cite{Jed:Chm:Szn:15}. However, it should be noted that
previously we could not speak about the energy or the temperature because only $q$ out of all neighbors were considered.

\section{Results for the quenched $q$-RRG} 
\begin{figure}[b!]
	\centerline{\epsfig{file=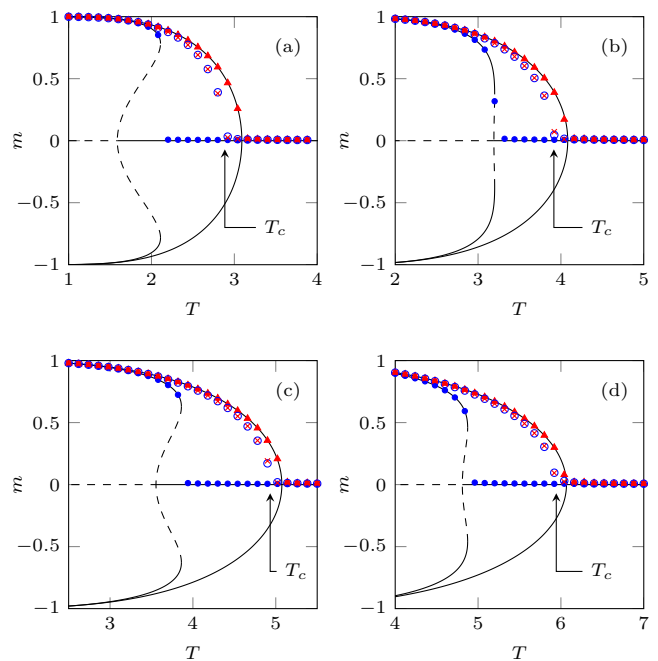}}
	\caption{\label{fig_quenched}Phase diagrams for the Ising model on annelad and quenched $q$-RRG with different degrees: (a) $q=4$, (b) $q=5$, (c) $q=6$, (d) $q=7$. Lines indicate analytical predictions of the mean-field approximation in the case of annealing: lines that are solid (stable states) for all values of $T$ correspond to heat-bath dynamics (HB annealed) and lines that are partially dashed (unstable states) to Metropolis (M annealed). Markers represent outcomes of Monte Carlo simulations: $\bullet$ M annealed, $\blacktriangle$ HB annealed, $\circ$ M quenched, $\times$ HB quenched. The critical temperature $T_c$ predicted by Eq.~(\ref{eq:crit_quenched}) almost exactly concurs with the simulated one on quenched graphs.}
\end{figure}
We investigate the model by Monte Carlo simulations and estimate the magnetization of the system as an ensemble average over $M$ samples:
\begin{equation}
m=\frac{1}{M}\sum_{j=1}^M m_j,
\end{equation}
where $m_j$ is the stationary magnetization in $j$-th sample, defined as:
\begin{equation}
m_j=\frac{1}{N}\sum_{i=1}^N S_i^j,
\end{equation}
here $S_i^j$ denotes a value of an $i$-th spin in the $j$-th sample in the stationary state. In the case of the quenched approach for each $j$-th sample we build a new random regular graph of size $N$, in which degrees of all nodes are equal $q$, but the network does not change in time. In the case of the annealed approach we change the network after each single update defined by the algorithm in the previous section.

For the quenched approach, we expect that Metropolis and heat-bath algorithms will give exactly the same results. Moreover, results should be consistent with the general analytical formula for the critical temperature $T_c$ for the Ising model with coupling constant $J$ on an ``equilibrium'' random network with the degree distribution $P(k)$, derived by  Dorogovtsev \textit{et al.} \cite{Dor:Gol:Men:02}:
\begin{equation}
\frac{J}{T_c}=\frac{1}{2} \ln \left( \frac{\langle k^2\rangle}{\langle k^2\rangle-2\langle k\rangle}\right).
\end{equation}
For $q$-RRG degrees of all nodes are equal to $q$. Moreover, we have assumed that $J=1$; therefore, the critical temperature should be equal:
\begin{equation}
T_c(q)=2 \left[ \ln \left( \frac{q}{q-2}\right)\right]^{-1}.
\label{eq:crit_quenched}
\end{equation}
Indeed, if we look at Fig. \ref{fig_quenched} we see that results for HB and M dynamics are the same, and the critical temperature is properly described by Eq. (\ref{eq:crit_quenched}), which is also clearly visible in the left panel of Fig. \ref{fig:Tc}.

\section{Results for the annealed $q$-RRG} 
On one hand, theoretical description simplifies significantly if we replace the quenched 
network by the annealed one because in such a case we can easily derive the transition probabilities for the system and write down the rate and/or the master equation, as it was done in the previous paper \cite{Jed:Chm:Szn:15}. On the other hand, the behavior on the annealed graph, at least for the Ising model with Metropolis dynamics, is much more complex including: (1) a switch from a continuous to a discontinuous phase transition at $q = 4$ and (2) an unexpected oscillatory behavior of the hysteresis, expanding for even values of $q$ and shrinking for odd values of $q$.

Here, following the reasoning presented in \cite{Jed:Chm:Szn:15}, we derive transition probabilities for the Ising model with HB dynamics on the annealed $q$-RRG. This allows us to calculate stationary magnetization as well as the critical temperature.
In a single update, three events are possible -- the number of spins `up': 
 \begin{enumerate}
 	\item 
 	increases by 1 ($N_{\uparrow} \rightarrow N_{\uparrow}+1$) and simultaneously $m \rightarrow m+2/N$ with probability $\gamma^+$,
 	\item 
 	decreases by 1 ($N_{\uparrow} \rightarrow N_{\uparrow}-1$) and simultaneously  $m \rightarrow m-2/N$ with probability $\gamma^-$,
 	\item
 	or remains constant with probability $1-\gamma^+ - \gamma^-$.
 \end{enumerate}
To simplify calculations we define a new macroscopic variable -- the concentration of `up' spins at time $t$:
\begin{equation}
c(t)=\frac{N_{\uparrow}(t)}{N}=\frac{m(t)+1}{2}.
\label{eq:rel_cm}
\end{equation}
Using this variable we can write down the transition probabilities $\gamma^+,\gamma^-$ for the infinite system:
\begin{eqnarray}
\gamma^+ (c) = \sum_{k=0}^{q} {q \choose k} c^{q-k}(1-c)^{k+1} \frac{1}{1+  e^{\frac{-2}{T} (q-2k) }} \nonumber, \\
\gamma^- (c) = \sum_{k=0}^{q} {q \choose k} (1-c)^{q-k}c^{k+1} \frac{1}{1+  e^{\frac{-2}{T} (q-2k) }}.
\label{eq:transit_prob}
\end{eqnarray}
In above equations we have replaced $c(t)$ by $c$ for brevity of description, but one should remember that $c$ evolves in time, and this evolution is described by the rate equation \cite{Kra:Red:Ben:10}:
\begin{eqnarray}
c(t+1) = c(t) +  \left[\gamma^+(c)-\gamma^-(c)\right],
\end{eqnarray}
where the unit time has been defined as $N$ elementary updates described by the algorithm in section II, as usually one Monte Carlo step is defined.

In the steady state $c(t+1) = c(t)=c$, which is equivalent to the condition:
\begin{eqnarray}
\gamma^+(c) - \gamma^-(c) = 0.
\label{F0}
\end{eqnarray}

\begin{figure}[!t]
	\centerline{\epsfig{file=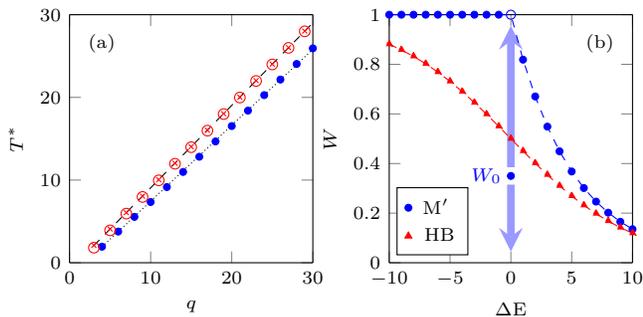}}
	\caption{\label{fig:criticalT} (a) Transition temperature as a function of $q$ for the kinetic Ising model on the annealed $q$-RRG with two dynamics: $\bullet$ Metropolis and $\times$ heat-bath. Markers $\circ$ represent the results obtained from Eq.~(\ref{eq:crit_quenched}) for the quenched structure.  Note that in the case of quenching, both dynamics give the same critical temperature. Moreover, HB algorithm reproduces roughly the same results for both types of graphs, whereas M algorithm gives different value of $T_c$ on the annealed and quenched networks. (b) Transition rates as functions of $\Delta E$ for heat-bath and modified Metropolis dynamics. Note that when $W_0=1$, the modified Metropolis algorithm coincides with the original version.}
	\label{fig:Tc}
\end{figure}

Now, we are ready to derive the formula for the critical temperature $T_c$.
Let us define a quantity which can be thought of as a net force acting on our system in the following way \cite{Nyc:Szn:Cis:12}:
\begin{equation}
F(c,T)=\gamma^+(c,T)-\gamma^-(c,T).
\label{eq:netForce}
\end{equation}
In such a case, the lines of equilibrium are determined by the condition
\begin{equation}
F(c,T)=0.
\end{equation}
Differentiating the above equation with respect to the concentration gives
\begin{equation}
\frac{\partial F(c,T)}{\partial c}+\frac{\partial F(c,T)}{\partial T}\frac{dT}{dc}=0.
\label{eq:diffF}
\end{equation}
Note that the critical point corresponds to the extremum of $T(c)$. Hence, the first derivative of the temperature over $c$ vanishes at $(c,T)=(1/2,T_c)$, that is to say,
\begin{equation}
\left.\frac{dT}{dc}\right|_{(\frac{1}{2},T_c)}=0.
\end{equation}
Therefore, evaluating Eq.~(\ref{eq:diffF}) at the critical point gives us the condition for its derivation
\begin{equation}
\left.\frac{\partial F(c,T)}{\partial c}\right|_{(\frac{1}{2},T_c)}=0.
\label{eq:condTc}
\end{equation}
Using formulas for the transition probabilities, we obtain the equation from which the critical temperature can be derived numerically
\begin{equation}
\label{eq:criticalT}\sum_{i=0}^{q}\binom{q}{i}(2i-q)\tanh\frac{2i-q}{T_c}=2^q.
\end{equation}
For $q=4$, our solution coincides with the result presented in \cite{Lip:Gon:Lip:15}. Transition temperature as a function of $q$ is presented in the left panel of Figure~\ref{fig:criticalT}. As seen, $T_c$ derived from Eq.~(\ref{eq:criticalT}) for an annealed $q$-RRG approaches the critical temperature derived by Dorogovtsev \textit{et al.} \cite{Dor:Gol:Men:02} in the case of a quenched graph given by Eq. (\ref{eq:crit_quenched}). Furthermore, it is seen in Fig. ~\ref{fig:criticalT} that Metropolis algorithm, considered already in \cite{Jed:Chm:Szn:15}, gives different value of $T_c$ for the annealed graph.

\section{Different dynamics}
Figure~\ref{fig_quenched} illustrates differences of the out-comes from two popular dynamics: Metropolis and heat-bath. Furthermore, it has been shown that the heat-bath algorithm reproduces almost the same results for annealed and quenched graphs. The difference  is only quantitative and decreases with the growing average degree $q$; see also Fig.~\ref{fig:criticalT} (left panel). On the other hand, Metropolis dynamics causes much more profound changes in the behavior of the system. On quenched networks only continuous phase transitions are observed whereas on annealed graphs both continuous and discontinuous transitions are present depending on the parameter $q$. Furthermore, the hysteresis exhibits oscillatory characteristics shrinking for odd values of $q$ and expanding for even ones \cite{Jed:Chm:Szn:15}. 

In order to comprehend why annealing for some dynamics causes once only quantitative and other qualitative differences, we introduce and investigate several new dynamics which fulfill the detailed balance condition and reproduce the Boltzmann distribution in equilibrium for the model placed on quenched $q$-RRG. In other words, the transition probabilities satisfy Eq.~(\ref{det_bal}). 
\begin{figure}[t!]
	\centerline{\epsfig{file=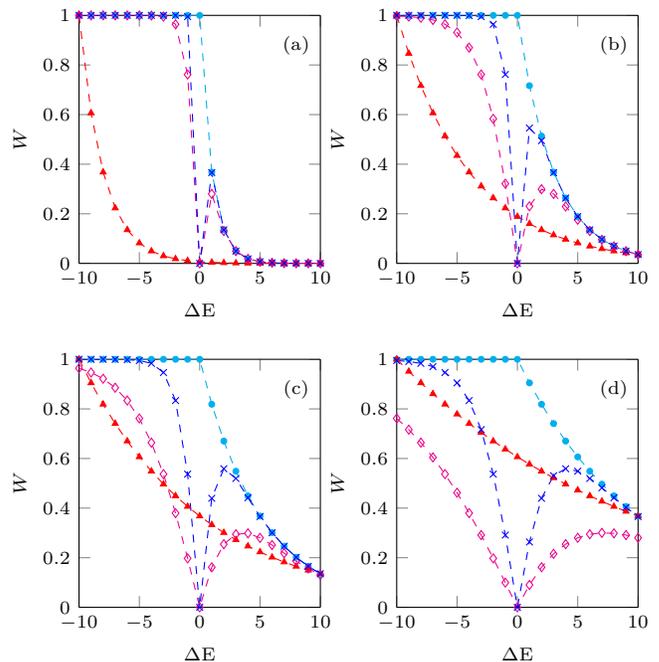}}
	\caption{\label{fig:differentDyn}Transition rates as functions of $\Delta E$ for different dynamics: $\bullet$ $W_{\text{M}}$, $\times$ $W_{\tanh}$ with $a=3$, {\scriptsize$\Diamond$} $W_{\tanh}$ with $a=1$, and $\blacktriangle$ $W_{\exp}$. Consecutive panels correspond to higher temperatures: (a) $T=1$, (b) $T=3$, (c) $T=5$, and (d) $T=10$.}
\end{figure}

The most natural choice for such a dynamics comes from the modification of the Metropolis algorithm:
\begin{equation}
\label{eq:modifiedM}
W_{\text{M}'}= \begin{cases}
\exp(-\beta\Delta E) &(\Delta E >0)\\
W_0 &(\Delta E=0)\\
1 &(\Delta E<0).
\end{cases}
\end{equation}
The above modification is identical with the original Metropolis dynamics, except of the case $\Delta E=0$. For M algorithm $W(\Delta E=0)=1$, on the other hand for the HB dynamics $W(\Delta E=0)=1/2$. Here we propose a~ generalized version in which $W(\Delta E=0)=W_0$, where $W_0 \in [0,1]$ is a parameter; see the right panel of Fig.~\ref{fig:criticalT}. Such a modification reminds of the generalized zero-temperature Glauber dynamics introduced in \cite{God:Luc:05}, which occurred to be particularly interesting from theoretical point of view \cite{Lip:99,Rad:Vil:Mey:07,Sko:Szn:Top:12,Spi:Kra:Red:01,Ole:Kra:Red:11}.  As we will see, the above dynamics, although artificial and not motivated by any physical processes, will also turn out to be particularly intriguing. Nevertheless, other transition rates that fulfill Eq.~(\ref{det_bal}) are also possible, and we will briefly present results for two of them:
\begin{equation}
	W_{\exp}=C\exp\left(-\frac{1}{2}\beta\Delta E\right),
\end{equation}
where $C=\exp(-\beta q)$ is a normalizing constant and
\begin{equation}
	W_{\tanh}= \min\left[1,\exp(-\beta\Delta E)\tanh\left|a\beta\Delta E\right|\right],
	\label{eq:Wtan}
\end{equation}
where $a$ is a parameter. Transition probabilities generated by the proposed dynamics are presented in Fig.~\ref{fig:differentDyn} for several temperatures. Note that for growing $a$ the above transition rate approaches $W_{\text{M}'}$ with $W_0=0$.

Although all dynamics satisfy detailed balance and therefore give the same results on quenched random $q$-regular graphs, the outcomes differ when we anneal the network.
For $W_{\exp}$ phase transition is continuous for all values of parameters, like in the case of quenching, but diagrams are shifted towards higher temperatures. Consequently, it results in the larger values of the critical points.

\begin{figure}[t!]
	\centerline{\epsfig{file=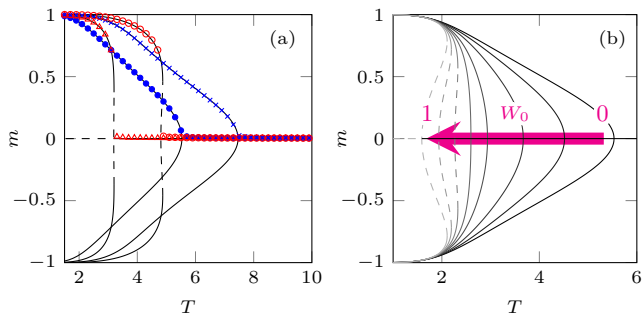}}
	\caption{\label{fig:modifiedM}(a) Phase diagrams for the Ising model with the modified Metropolis algorithm Eq.~(\ref{eq:modifiedM}) on annealed $q$-RRG for $W_0=0$. Markers represent outcomes of Monte Carlo simulations: $\bullet$ $q=4$, $\vartriangle$ $q=5$, $\times$ $q=6$, and $\circ$ $q=7$; Lines indicate analytical predictions of the mean-field approximation. Solid ones correspond to stable states, and dashed ones refer to unstable points. (b) Phase diagrams for the Ising model with $\text{M}'$ dynamics and $q=4$ for different values of $W_0$. Lighter lines correspond to higher $W_0$.  }
\end{figure}

More thought provoking behavior of the model is induced by the modified Metropolis algorithm $W_{\text{M}'}$ with $W_0=0$. Then the system exhibits two types of phase transitions. For odd values of $q$ both Metropolis dynamics, modified and original one, are consistent and produce qualitatively the same results, in a sense that all transitions for $q>3$ are discontinuous. However, for even values of $q$ all transitions become continuous; see the left panel of Fig.~\ref{fig:modifiedM}. It should be recalled here that for $W_0=1$, which corresponds to original Metropolis results were almost opposite -- though phase transitions were discontinuous for all $q>3$, yet just for even values of $q$ discontinuity was stronger (larger jump of the order parameter and hysteresis). 

\begin{figure}[t!]
	\centerline{\epsfig{file=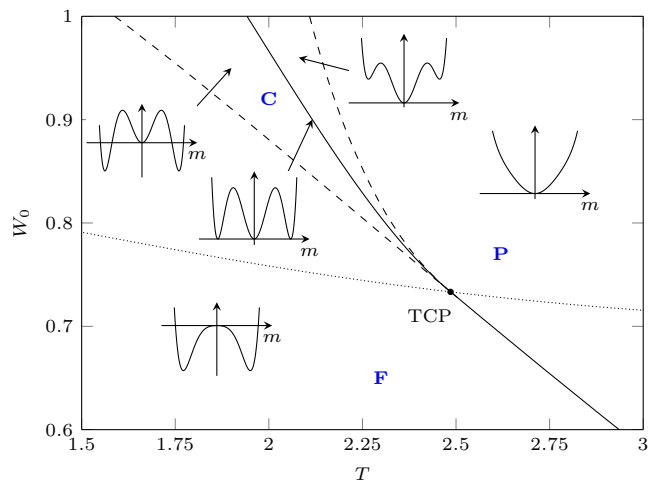}}
	\caption{\label{fig:phaseReal} Phase diagram for the non-equilibrium kinetic Ising model on the annealed $q$-RRG with the modified Metropolis dynamics for $q=4$. Along the solid line, a phase transition occurs: below the tricritical point (TCP) there is a continous phase transition and above TCP the continuous transition line splits up into two spinodal (dashed) lines, where the transition becomes first order. Along the dotted line, the term with $m^4$ in the Landau potential vanishes. The diagram divides a phase space into three distinctive areas: F where ferromagnetic (ordered) phase is present, P with paramagnetic (disordered) phase, and C a region where both phases coexist. Additionally, a schematic Landau potential is sketched for each area.}
\end{figure}

The right panel of Fig.~\ref{fig:modifiedM} presents how the phase diagram changes under the influence of $W_0$ for the model with $q=4$. It is seen that there is a critical value of $W_0$ for which the transition type for even values of $q$ switches from discontinuous to continuous one. We can derive this critical point $W^*_0$ as well as categorize phase transitions based on the classical Landau theory \cite{Lan:37,Nyc:Szn:Cis:12} by introducing an effective potential $V(m)$ and studying its behavior for small values of the order parameter $m$, i.e., near zero.
Having the net force $F(m,T)$, from Eqs.~(\ref{eq:netForce}) and (\ref{eq:rel_cm}), we can construct the potential in the following way:
\begin{equation}
V(m,T)=-\int F(m,T)dm.
\label{eq:potential}
\end{equation}
In the vicinity of zero, we can expand it into a Taylor series, and neglect terms of the higher-order than $m^6$. This procedure leads to the potential in the form:
\begin{equation}
V(m)=Am^2+Bm^4+Cm^6+O(m^8),
\label{eq:potentialTaylor}
\end{equation}
where coefficients $A$, $B$, and $C$ depend on the model and can be derived directly from Eq.~(\ref{eq:potential}) as:
\begin{align}
A&=-\left.\frac{1}{2!} \partial_m F(m,T) \right|_{(0,T)},\\
B&=-\left.\frac{1}{4!} \partial^3_mF(m,T) \right|_{(0,T)},\\
C&=-\left.\frac{1}{6!} \partial^5_mF(m,T) \right|_{(0,T)}.
\end{align}
When $A>0$, then the second derivative of Eq.~(\ref{eq:potentialTaylor}) near $m=0$ is positive, and at this point $V(m)$ has a minimum; thus, a disordered phase is stable. On the other hand, when $A$ is negative, a disordered phase is unstable. Therefore, a condition $A=0$ allows us to derive the boundary of a region where a paramagnetic phase is present; note that this is equivalent to Eq.(\ref{eq:condTc}).
The order of a transition is determined by the coefficient $B$. If it is positive, the system undergoes continuous phase transition 
. When it is negative, a discontinuous transition appears, and there is an area, bounded by the spinodal lines (dashed lines in Fig.~\ref{fig:phaseReal}), where two phases may coexist. Within the modified Metropolis dynamics, defined by Eq.(\ref{eq:modifiedM}), there is a ferromagnetic-paramagnetic continuous phase transition for small values of $W_0$ and above a certain tricritical point (TCP) the continuous transition line splits up into two spinodal lines, where the transition becomes first order (see Fig.~\ref{fig:phaseReal}). Similar behavior has been observed recently for the non-equilibrium Ising model with two heat-baths, but in that case the temperature of the heat-bath for links $T_L$ has driven the tricritical behavior \cite{Par:Hoh:17}.

Because TCP is  a place where all three sections: the coexistence area and regions with purely ordered and disordered phases meet, it might be determined by fulfilling simultaneously two conditions $A=0$ and $B=0$. Moreover, note that TCP settles the value of $W_0^*$. 

\begin{figure}[t!]
	\centerline{\epsfig{file=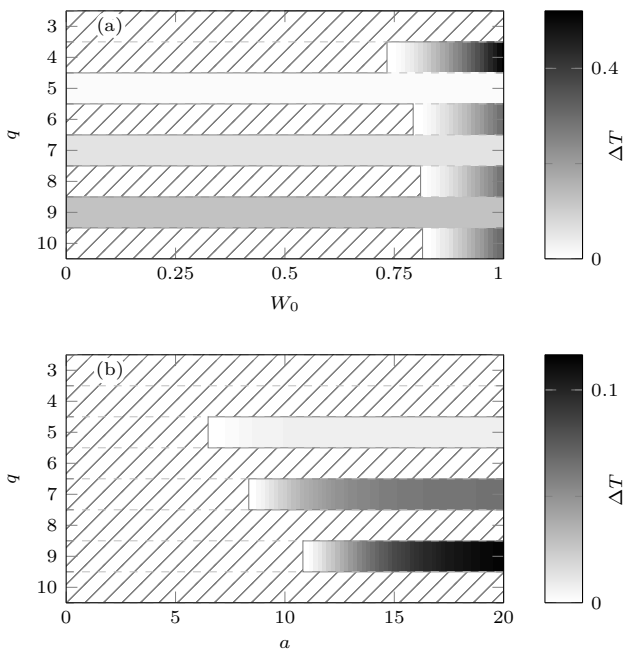}}
	\caption{ 
		The dependence between the type of the phase transition and model's parameters for (a) the modified Metropolis dynamics and (b) the dynamics with $W_{\tanh}$ rates. There are two characteristic regimes: with continuous (marked by oblique lines) and discontinuous phase transitions. In the case of discontinuous transitions, darker areas indicate wider hysteresis $\Delta T$, measured as the difference between the upper and the lower spinodal lines. Note that in (a) the vertical section along $W_0=1$ corresponds to the original Metropolis dynamics, studied in \cite{Jed:Chm:Szn:15}. For even values of interacting neighbors $q$, there exists a critical value of $W_0$ where discontinuous phase transition becomes continuous. Similar behavior is observed for the model with $W_{\tanh}$ transition probabilities for odd $q$ values (b), but this time below a certain level of $a$, all transitions convert into continuous ones.}
	\label{fig:phaseSpaceModM}
\end{figure}

The entire phase diagram for the model with the modified Metropolis dynamics for q=4 is illustrated in Fig.~\ref{fig:phaseReal}.
The analogical behavior is observed for grater even values of interacting neighbors.
In this particular case ($q=4$), the critical point lies at $W_0^*=11/15$, but in general its value depends on $q$.
The dependence between the type of the phase transition and model's parameters $q$ and $W_0$ for the modified Metropolis dynamics
is presented in the top panel of Fig.~\ref{fig:phaseSpaceModM}. For discontinuous transitions, shading indicates the width of the hysteresis; darker areas correspond to wider metastable region $\Delta T$ where two phases may coexist. Oblique lines indicates the presence of continuous phase transitions. The boundary between these two transition types for a given $q$ corresponds to the tricritical point.

Based on the above result, we can draw a conclusion that by altering transition probabilities at $\Delta E=0$ one can control the nature of the phase transition. Moreover, lower flipping rates are connected with continuous transitions. 
In order to check whether indeed only the central point (i.e. $\Delta E=0$) of the dynamics has this profound impact on the model behavior we investigate the system with $W_{\tanh}$ rates, defined by Eq. (\ref{eq:Wtan}). In this case, parameter $a$ does not affect $W(\Delta E=0)$, and for all conditions $W(\Delta E=0)=0$. Nonetheless, its higher value increases in general the flipping probabilities, which is clearly seen in Fig.~\ref{fig:differentDyn}. Note that large $a$ reproduces the outcome obtained for the model drove by the modified Metropolis algorithm with $W_0=0$ since then both dynamics coincide; look at Fig.~\ref{fig:phaseSpaceModM} and compare the vertical section along $W_0=0$ in the top panel with the section along $a=20$ in the bottom one. However, there is a certain value of the parameter $a$ below which all transitions become continuous, like for $W_{\text{HB}}$ or $W_{\exp}$ dynamics.

\section{Conclusions}
Ising model, although already 100 years old still inspires many researchers and raises new questions \cite{Rad:Vil:Mey:07,Sko:Szn:Top:12,Spi:Kra:Red:01,Ole:Kra:Red:11,Roy:Bis:Sen:14,Roy:Sen:15,Men:Odo:00}. Recently we have introduced, under the name the $q$-neighbor Ising model, a seemingly small modification of the kinetic Ising model with Metropolis dynamics allowing each spin to interact only with $q$ neighbors \cite{Jed:Chm:Szn:15}. Surprisingly, this modification leads to a switch from a continuous to a discontinuous phase transition at $q = 4$, and it causes an unexpected oscillating behavior of the hysteresis for $q \ge 4$ -- it expands for even values of $q$ and shrinks for odd values of $q$. This was very puzzling phenomenon because extensive studies have shown not only that a phase transition in the equilibrium Ising model is continuous but also is characterized with the same set of bulk critical exponents on quenched and annealed networks \cite{Lee:09}.

However, recently it has been shown that  the $q$-neighbor Ising model is a limiting case of a non-equilibrium system with two 
heat baths -- one for the Ising spins at temperature $T_s$ and second for links of the graph at temperature $T_L$ \cite{Par:Hoh:17}. For $T_L=\infty$ such a generalized model reduces to the $q$-neighbor Ising model investigated in \cite{Jed:Chm:Szn:15}. This means that the $q$-neighbor Ising model is out of equilibrium on contrary to the Ising model on the annealed graph considered in \cite{Lee:09}. However, the question arises whether just being out of equilibrium causes tricritical behavior, i.e., switch from continuous to discontinuous phase transition. Maybe such an exotic behavior is observed exclusively for Metropolis algorithm? Furthermore, what is the role of dynamics for non-equilibrium systems? 

Referring above questions, we have reformulated the $q$-neighbor Ising model in terms of the classical Ising model on the $q$-RRG and investigated it under several different dynamics belonging to the broad class of the generalized single-spin flip Glauber dynamics \cite{Glauber,God:Luc:05}. On the quenched graph, which within the generalized model proposed in \cite{Par:Hoh:17} corresponds to the zero temperature of links ($T_L=0$), all dynamics fulfill the detailed balance condition and lead to the equilibrium Boltzmann distribution. Therefore, it is not surprising that they all give exactly the same results and reproduce analytical formulas for the critical temperature derived by Dorogovtsev \textit{et al.} \cite{Dor:Gol:Men:02}. 

On the annealed non-equilibrium network, which within the generalized model proposed in \cite{Par:Hoh:17} corresponds to the infinite temperature of links ($T_L=\infty$), each dynamics leads to completely different results. For example, the 
model with heat-bath algorithm on the annealed network displays always a continuous phase transition, and the critical temperature only slightly deviates from the equilibrium one. On the other hand, the generalized Metropolis algorithm gives qualitatively different behavior, which depends on the model's parameter $W_0$. For $W_0=1$, we obtain the original Metropolis dynamics, which means that the transition switches to discontinuous at $q=4$, and hysteresis exhibits oscillatory behavior -- expanding for even values of $q$ and shrinking for odd values of $q$. Yet, for $W_0<W_0^*(q)$, the behavior is even more exotic -- phase transition switches alternately from continuous to discontinuous phase transition and back. Additionally, we have investigated two other types of dynamics, and it occurred that $W_{\exp}$ displays always a continuous phase transition while $W_{\tanh}$ switches between two types of the phase transitions. We have observed that generally higher flipping probabilities lead to discontinuous phase transitions.

We could of course propose many other dynamics which fulfill the detailed balance condition on the quenched graph, but it was not our aim. We asked the question about the role of dynamics for non-equilibrium models. On one hand, there is no excuse of referring to different dynamics or algorithms if the model is out of equilibrium, because in such a case there is no condition, such as detailed balance, which guarantees the unique steady state. On the other hand, referring to the particular dynamics within non-equilibrium models still happens \cite{Lip:Gon:Lip:15,Rad:Vil:Mey:07}. Such practices are not so much surprising, because it is quite common that methods of equilibrium statistical physics are transferred to the field of non-equilibrium systems and they give often reasonable results \cite{Hen:Hin:Lue:08}. Thus, one could have an intuition that the choice of a particular dynamics for non-equilibrium system will not influence significantly the outcome of the model. 

Because results obtained in \cite{Jed:Chm:Szn:15} have shown that such an intuition may be very wrong, we have decided to examine several different dynamics and check how they would influence results in two limiting cases: quenched ($T_L=0$) and annealed ($T_L=\infty$). We have shown that indeed some dynamics (including heat-bath) are more resistant to the rewriring a network and display always continuous phase transition, whereas other (including Metropolis) are much more fragile and rewriring a network changes the type of the phase transition. Furthermore, our results suggest that we can generally expect discontinuous phase transitions for higher flipping rates $W_0$, i.e., in fact for larger noise, which is generally interesting result and coincides with observations made in \cite{Par:Hoh:17}, where discontinuity has been driven by higher temperature of a network. Of course there are more possibilities to to incorporate noise to the system. For example one could investigate the model on temporal network and introduce rewriring time $\tau$, as a parameter. Because $\tau=\infty$ corresponds to $T_L=0$ (i.e., quenched network) and $\tau=1/N$ to $T_L=\infty$ (i.e., annealed network), one could expect that there is a critical time scale $\tau^*$ that separates these two regimes. Indeed preliminary studies confirm these predictions \cite{Mar:Szn:17}.

It is known that switch from continuous to discontinuous phase transition can be caused by many factors. In the equilibrium statistical physics one of the best known examples is the regime-switch in the Potts model (for review see \cite{Wu:82}): for $q>4$ the model undergoes a first-order phase transition, whereas a  second-order phase transition for smaller values of $q$, with $q$ being the number of states of the spin. Moreover, it is common that systems exhibiting a discontinuous phase transition in high space dimensions may display a continuous transition below a certain critical dimension \cite{Hen:Hin:Lue:08}. For non-equilibrium systems it has been shown that by increasing the number of interacting neighbors, the fluctuations are diminished and the transitions become sharper or even change their type to discontinuous \cite{Nyc:Szn:Cis:12,Jed:Chm:Szn:15,Odo:Szo:96}. Yet, recent research has shown that for non-equilibrium systems switch from continuous to discontinuous phase transition and related tricriticality may be also caused by a high temperature of a network \cite{Par:Hoh:17} or larger flipping probabilities. These results suggest that incorporating certain type of noise to the non-equilibrium system can be responsible for tricriticality. Yet, this hypothesis requires further research.

However, the main outcome of this paper is the answer to the question posed in the introduction.  Our aim was to check if being out of equilibrium, one can still speak about the Ising model with certain dynamics. As we have shown, each out of equilibrium dynamics leads to completely different behavior -- some dynamics are more fragile for the non-equilibrium driving than others. In fact, it is even hard to speak about the dynamics or algorithm in case of such a non-equilibrium system because dynamics is an immanent feature of the model.

\begin{acknowledgments}
This work was supported by funds from the National Science Centre (NCN, Poland) through grants no. 2015/18/E/ST2/00560 (to AC) and no. 2016/21/B/HS6/01256 (to AJ and KSW).
\end{acknowledgments}


\end{document}